\documentclass[12pt,a4paper,epsf]{article}
\usepackage{graphics}
\usepackage{amssymb,amsmath}
\usepackage[dvips]{lscape,graphicx}
\usepackage{cite}
\usepackage{longtable}

\newcommand{\ct}{\cite}
\newcommand{\lb}{\label}

\newcommand{\bc}{\begin{center}}
\newcommand{\ec}{\end{center}}
\newcommand{\bd}{\begin{displaymath}}
\newcommand{\ed}{\end{displaymath}}
\newcommand{\be}{\begin{equation}}
\newcommand{\ee}{\end{equation}}
\newcommand{\ba}{\begin{array}}
\newcommand{\ea}{\end{array}}
\newcommand{\bea}{\begin{eqnarray}}
\newcommand{\eea}{\end{eqnarray}}
\newcommand{\bt}{\begin{tabular}}
\newcommand{\et}{\end{tabular}}

\newcommand{\bp}{\begin{picture}}
\newcommand{\ep}{\end{picture}}
\newcommand{\bfi}{\begin{figure}}
\newcommand{\efi}{\end{figure}}

\def\fun#1#2{\lower3.6pt\vbox{\baselineskip0pt\lineskip.9pt
\ialign{$\mathsurround=0pt#1\hfil##\hfil$\crcr#2\crcr\sim\crcr}}}

\begin{document}



\vspace{1cm}

\title{\LARGE \bf {The relation between the model of a crystal with defects
and Plebanski's theory of gravity}}
\author{\large D.L.~Bennett${}^{1}$
\footnote{dlbennett99@gmail.com}\,\,, C.R.~Das${}^{2}$\footnote
{crdas@cftp.ist.utl.pt}\,\,, L.V.~Laperashvili ${}^{3,}$${}^{4}$
\footnote{laper@itep.ru}\,\, and H.B.~Nielsen${}^{4}$
\footnote{hbech@nbi.dk}\\[5mm]
\itshape{\large ${}^{1}$ Brookes
Institute for Advanced Studies,
Copenhagen, Denmark}\\
\itshape{\large ${}^{2}$ Centre for Theoretical Particle Physics,
Lisbon, Portugal}\\
\itshape{\large ${}^{3}$The Institute of Theoretical and
Experimental Physics, Moscow, Russia}\\
\itshape{\large ${}^{4}$ The Niels Bohr Institute, Copenhagen,
Denmark}}

\date{}

\maketitle

\thispagestyle{empty}

\begin{abstract}

In the present investigation we show that there exists a close
analogy of geometry of spacetime in GR with a structure of defects
in a crystal. We present the relation between the Kleinert's model
of a crystal with defects and Plebanski's theory of gravity. We
have considered the translational defects -- dislocations, and the
rotational defects -- disclinations -- in the 3- and 4-dimensional
crystals. The 4-dimensional crystalline defects present the
Riemann-Cartan spacetime which has an additional geometric
property - "torsion" -- connected with dislocations. The world
crystal is a model for the gravitation which has a new type of
gauge symmetry: the Einstein's gravitation has a zero torsion as a
special gauge, while a zero connection is another equivalent gauge
with nonzero torsion which corresponds to the Einstein's theory of
"teleparallelism". Any intermediate choice of the gauge with
nonzero connection $A^{IJ}_{\mu}$ is also allowed. In the present
investigation we show that in the Plebanski formulation the phase
of gravity with torsion is equivalent to the ordinary or
topological gravity, and we can exclude a torsion as a separate
dynamical variable.

\end{abstract}

\thispagestyle{empty}

\clearpage\newpage

\section{Introduction}

There exists a close analogy of geometry of spacetime in General
Relativity (GR) with a structure of defects in a crystal
\ct{1,2,3}. The crystal's defects present a special version of the
curved Riemannian spacetime - the Riemann-Cartan spacetime which
has an additional geometric property called "torsion" \ct{4}. In
the absence of matter, the world crystal is a model for Einstein's
gravitation with a new type of gauge symmetry in which a zero
torsion is a special gauge, while a zero connection (and the
absence of the Cartan curvature) is another equivalent gauge
leading to the Einstein's theory of "teleparallelism" \ct{5}. The
relation of torsion to electromagnetism \ct{6} makes photon
massive. If torsion cannot be seen experimentally, then photon is
massless. Limits on the range of magnetic fields emerging from
planets and stars give the upper bounds on the photon mass which
is extremely small: $m_{\gamma} < 10^{-27}$ eV (e.g. $m_{\gamma}<
10^{-60}$ g). Laboratory experiments yield much weaker bound:
$m_{\gamma} < 3\times 10^{-16}$ eV (e.g. $m_{\gamma} < 10^{-49}$
g). The recent critical discussion can be found in Ref.~\ct{7}.
Later in Refs.~\ct{8,9,10} the relationship between torsion and
the spin density of the gravitational field was considered.

\section{The structure of defects in a crystal}

\subsection{The structure of defects in the 3-dimensional crystal}

Crystalline defects may be generated via so called "Volterra
process", when layers or sections of matter are cut from a crystal
with a subsequent smooth rejoining of the cutting surfaces.

A crystal can have two different types of topological defects
\ct{1,2,3}, which are line-like defects in the 3-dimensional space
and form world surfaces in the 4-dimensional crystal (they may be
objects of string theory).

A first type of topological defects are translational defects
called {\it dislocations}: a single-atom layer is removed from the
crystal and the remaining atoms relax to equilibrium under the
elastic forces (see Fig.~1).

\begin{figure} [htb]
\centering
\includegraphics[height=50mm,keepaspectratio=true,angle=0]{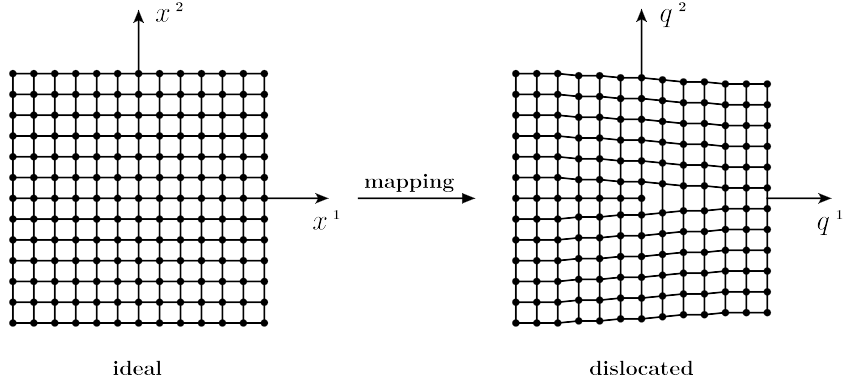}
\caption{ \it Formation of a dislocation line in a crystal.}
\end{figure}

A second type of defects is of the rotation type and called {\it
disclinations}. They arise by removing an entire wedge from the
crystal and re-gluing the free surfaces (see Fig.~2).

\begin{figure} [htb]
\centering
\includegraphics[height=45mm,keepaspectratio=true,angle=0]{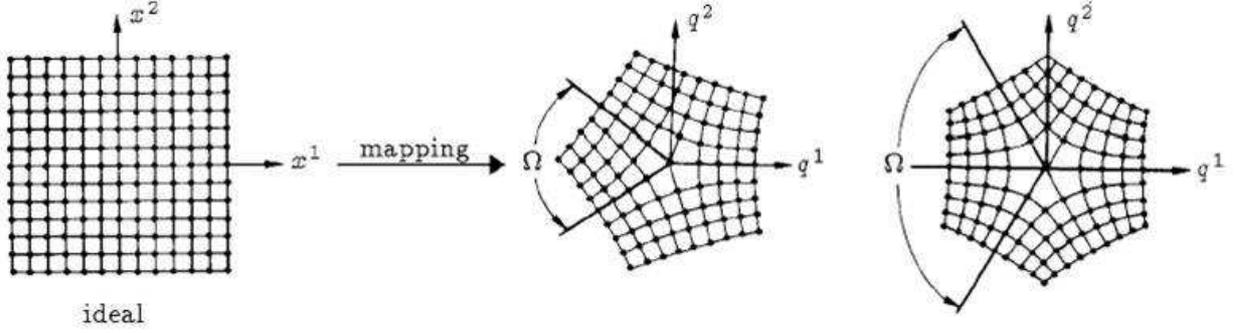}
\caption{ \it Formation of disclinations in a crystal.}
\end{figure}

Considering the displacement field $u_i(\vec{x})$ in the
3-dimensional crystal, we can calculate the dislocation density
given by the following tensor:
\be \alpha_{ij} =
\epsilon_{ikl}\bigtriangledown_k\bigtriangledown_lu_j(\vec x).
                              \lb{1}
\ee
The local rotation field
\be
\omega_i = \frac 12 \epsilon_{ijk}[\bigtriangledown_ju_k(\vec
x) - \bigtriangledown_ku_j(\vec x)]      \lb{2} \ee
determines the disclination density:
\be
     \Theta_{ij} = \epsilon_{ikl}\bigtriangledown_k
     \bigtriangledown_l\omega_j(\vec x).
                              \lb{3} \ee
The defect densities satisfy the conservation laws:
\be \bigtriangledown _i\Theta_{ij} = 0, \quad
\bigtriangledown_i\alpha_{ij} = - \epsilon_{jkl} \Theta_{kl}.
                              \lb{4} \ee
The plastic displacement field $u_i^p(\vec{x})$ doesn't obey the
integrability conditions of Schwarz, i.e. near the plastic
deformations the second derivatives don't commute:
\be
   (\bigtriangledown_i\bigtriangledown_j -
    \bigtriangledown_j\bigtriangledown_i)u_k^p(\vec x) \neq 0.
                                      \lb{5} \ee
Eqs.~(\ref{4}) are Bianchi identities if densities $\Theta_{ij}$
and $\alpha_{ij}$ are expressed in terms of the plastic gauge
fields $\beta_{ij}^p$ and $\phi_{ij}^p$ \ct{3}:
\be   \beta_{ij}^p \equiv \bigtriangledown_iu_j^p(\vec x)
        \lb{6} \ee
- for dislocation, and
\be   \phi_{ij}^p \equiv \bigtriangledown_i\omega_j^p(\vec x)
        \lb{7} \ee
- for disclination. Then the defect densities
\be
    \Theta_{ij} = \epsilon_{ikl}\bigtriangledown_k\phi_{lj}^p
    \quad {\mbox{and}}\quad \alpha_{ij} =
    \epsilon_{ikl}\bigtriangledown_k\beta_{lj}^p + \delta_{ij}
    \phi_{kk}^p - \phi_{ji}^p
                                     \lb{8} \ee
are invariant under the following gauge transformations:
\be \beta_{ij}^p \to \beta_{ij}^p + \bigtriangledown_iu_j^p(\vec
x) - \epsilon_{ijk}\omega_k^p(\vec x)\quad {\mbox{and}}\quad
 \phi_{ij}^p \to  \phi_{ij}^p +  \bigtriangledown_i\omega_j^p.
                                     \lb{9} \ee
Here we have chosen to use the same symbols for the two gauge
functions $u_k^p(\vec{x})$ and $\omega_k^p(\vec{x})$ as we used
for the fields describing the state of the (Kleinert) crystal,
because these quantities have the same character. But it is of
course a different meaning to a gauge function that can be chosen
arbitrarily and the true field describing the crystal. Therefore,
\be
     h_{ij} \equiv \beta_{ij}^p + \epsilon_{ijk}\omega_k^p
              \quad {\mbox{and}}\quad
      A_{ijk} \equiv \phi_{il}^p\epsilon_{ljk}
                                   \lb{10} \ee
are {\it translational} and {\it rotational} defect gauge fields
in the crystal.

In the Riemann-Cartan geometry of the 3-dimensional crystal the
{\it connection} is:
\be   \Gamma_{ijk} = \bigtriangledown_i\bigtriangledown_ju_k^p,
                         \lb{11} \ee
the {\it torsion} is:
\be   S_{ijk} = \frac 12(\Gamma_{ijk} - \Gamma_{jik}),
                                 \lb{12} \ee
and the {\it curvature} is:
\be  R_{ijkl} = (\bigtriangledown_i\bigtriangledown_j -
    \bigtriangledown_j\bigtriangledown_i)\bigtriangledown_ku_l^p
     =  \bigtriangledown_i\Gamma_{jkl} - \bigtriangledown_j\Gamma_{ikl}.
                                    \lb{13} \ee
In the 3-dimensional crystal the disclination density represents
the Einstein tensor $G_{ij}$ associated with the curvature
(\ref{13}):
\be G_{ij} = \Theta_{ij}(\vec x),
                     \lb{14} \ee
and the dislocation density represents the torsion:
\be     \alpha_{ij} = \epsilon_{ikl}S_{lkj}(\vec x).
                                    \lb{15} \ee

\subsection{Topological defects in the 4-dimensional crystal}

The geometry of the 4-dimensional Riemann-Cartan spacetime is
described by the direct generalizations of the translational and
rotational defect gauge fields $h_{ij}$ and $A_{ijk}$ on the
vierbein field $\theta_{\mu}^I$ and the spin connection
$A_{\mu}^{IJ}$ with $I,J=(0,1,2,3)$ and $I=(0,i)$ where $i=1,2,3$
is the spatial index.

In four dimensions the indices $I$ and $J$ which function as the
``flat'' index on the vierbein may be chosen to be identified with
the directions in the Kleinert-lattice (= world-chrystal) meaning
the index $k$ on the $u_k^p(\vec{x})$, so that we may call it now
- in four dimensions - $I$ instead of $k$, and write:
\be \theta^I_{\mu} = \partial_{\mu}u^p_I(x).  \lb{15a}  \ee
Similarly and analogously to the $A_{ijk}$ of equation (\ref{10})
we can in four dimensions write:
\be A^{IJ}_{\mu} = \partial_{\mu}\omega_{IJ}. \lb{15b}\ee
Here we have analogously to (\ref{2}) defined
\be \omega_{I J} =\nabla_I u_J^p(x)- \nabla_J u_I^p(x). \lb{15c}
\ee
In the traditional literature on gravity with spinning particles,
a special role is played by single-valued vierbein fields
$h^{\alpha}_{\mu}(x)$.

Considering the coordinate transformation:
\be   x^{\alpha} = x^{\alpha}(x^{\mu}),
                           \lb{16} \ee
we go from rectilinear coordinates $x^{\alpha}$ with
$\alpha=0,1,2,3$ to arbitrary curvilinear ones $x^{\mu}$ with
$\mu=0,1,2,3$, therefore, from the flat Minkowski metric
$\eta_{\alpha\beta}=(1,-1,-1,-1)$ to the induced metric:
\be g_{\mu\nu} = \eta_{\alpha\beta}
h_{\mu}^{\alpha}h_{\nu}^{\beta},
             \lb{17} \ee
where
\be
 h_{\mu}^{\alpha}\equiv \frac{\partial x^{\alpha}}{\partial x^{\mu}}
                        \lb{18} \ee
are the vierbein fields (tetrads). They define local coordinate
differentials $dx^{\alpha}$ by a transformation:
\be
 dx^{\alpha} = h^{\alpha}_{\mu}(x)dx^{\mu}.
                                    \lb{18a} \ee
Mathematically, the cutting and joining of the matter in the world
crystal may be described by {\it nonholonomic mappings} \ct{2} of
the next-neighbor atomic distance vectors. This mapping is not
integrable to a global coordinate transformation (\ref{16}). It is
described by a local transformation:
\be     dx^I=\theta^I_{\mu}(x)dx^{\mu},
                        \lb{19} \ee
whose coefficients  $\theta^I_{\mu}(x)$ have a nonvanishing curl:
\be  \partial_{\mu}\theta^I_{\nu} -
\partial_{\nu}\theta^I_{\mu}\neq 0,
                        \lb{20} \ee
what means that any coordinate transformation $x^I = x^I(x^{\mu})$
has to disobey the integrability conditions of Schwarz, and gives
the noncommutativity:
\be   (\partial_{\mu}\partial_{\nu} -
\partial_{\nu}\partial_{\mu})x^I(x_{\mu}))\neq 0.  \lb{21} \ee
Therefore, now the functions $x^I(x^{\mu})$ are {\it multivalued}
\ct{2}. This is in contrast to the curl of the usual vierbein
fields $h^{\alpha}_{\mu}(x)$ which is equal to zero and determines
purely Riemannian spacetime, i.e. spacetime without torsion.

A set of new nonholonomic coordinates $dx^I$ are related to
$dx^{\alpha}$ by a multivalued Lorentz transformation:
\be
 dx^I = \Lambda^I_{\alpha}(x)dx^{\alpha},
                              \lb{21a} \ee
and they are related to the physical $dx^{\mu}$ by the multivalued
tetrad fields:
\be dx^I = \theta^I_{\mu}(x)dx^{\mu}\equiv
\Lambda^I_{\alpha}(x)h^{\alpha}_{\mu}(x)dx^{\mu}.
                            \lb{21b} \ee
This procedure transforms the physical laws from the flat space to
spaces with curvature and torsion.

The condition (\ref{21}) is not enough to describe all topological
defects in a crystal. Also the coefficients $\theta^I_{\mu}$ in
Eq.~(\ref{19}) themselves must violate the Schwarz conditions
leading to the noncommutative derivatives:
\be
  (\partial_{\mu}\partial_{\nu} -
\partial_{\nu}\partial_{\mu})\theta^I_{\lambda}\neq 0.
 \lb{22} \ee
They are called {\it maltivalued tetrads} \ct{2}. The
multivaluedness distinguishes them from the well-known
single-valued tetrads (or vierbeins) $e_{\mu}^I\equiv h_{\mu}^I$
considered in the standard literature on gravity.\\
The field strength of $A_{\mu}^{IJ}$ is:
\be
     F_{\mu\nu}^{IJ} = \partial_{\mu}A_{\nu}^{IJ} -
         \partial_{\nu}A_{\mu}^{IJ} - [A_{\mu}, A_{\nu}]^{IJ},
                               \lb{23} \ee
and determines the Riemann-Cartan curvature \ct{4}:
\be
       R_{\mu\nu\lambda\kappa} =
 \theta_{\lambda}^I \theta_{\kappa}^J
 F_{\mu\nu}^{IJ}.
                  \lb{24} \ee
The field strength of $\theta_{\mu}^I$ is the torsion \ct{4}:
\be
     S_{\mu\nu}^I = D_{\mu}^{IJ}\theta_{\nu}^J - D_{\nu}^{IJ}\theta_{\mu}^J ,
                                  \lb{25} \ee
where
\be   D_{\mu}^{IJ} = \delta^{IJ}\partial_{\mu} - A_{\mu}^{IJ}
                             \lb{26} \ee
is a covariant derivative.

\section{Plebanski formulation of gravity}

The translational and rotational crystalline defect gauge fields
-- the tetrads $\theta_{\mu}^I$ (dislocations) and the spin
connection $A_{\mu}^{IJ}$ with $I,J=(0,1,2,3)$ (disclinations) --
were used by J.~Plebanski \ct{1a} for the construction of the
gravitational action. The main idea of Plebanski's formulation of
the 4-dimensional theory of gravity was to present the
gravitational action without metrics as a product of two 2-forms
\ct{1a,2a,3a,4a,5a,6a,7a}. These 2-forms were constructed using
the connection $A^{IJ}$ and tetrads $\theta^I$ as independent
dynamical variables.  The tetrads $\theta^I$ were used instead of
the metric $g_{\mu\nu}$.

Both $A^{IJ}$ and $\theta^I$ are 1-forms:
\be   A^{IJ} = A_{\mu}^{IJ}dx^{\mu}  \quad {\mbox{and}}\quad
       \theta^I = \theta_{\mu}^Idx^{\mu}.
                           \lb{1r} \ee
In Eq.~(\ref{1}) the indices $I,J = 0,1,2,3$ refer to the flat
space-time with Minkowski metric $\eta^{IJ} = {\rm diag}(1, -1,
-1, -1)$, which is tangential to the curved space with the metric
$g_{\mu\nu}$. The world interval is represented as
\be ds^2 =  \eta_{IJ}\theta^I \otimes \theta^J,  \lb{3r} \ee
i.e.
\be g_{\mu\nu} = \eta_{IJ} \theta^I_{\mu}\otimes \theta^J_{\nu}.
 \lb{4r} \ee
In the Plebanski's BF-theory \ct{1a}, the gravitational action
with nonzero cosmological constant $\Lambda$ is presented by the
integral:
\be  I_{GR} = \frac{1}{\kappa^2}\int
\epsilon^{IJKL}\left(B^{IJ}\wedge
    F^{KL} + \frac{\Lambda}{4}B^{IJ}\wedge B^{KL}\right),
                                  \lb{5r} \ee
where $\kappa^2=8\pi G$ ($G$ is the gravitational constant).

Below we use units $\kappa=1$.

The topological sector of gravity is described by the following
integral \ct{8a}:
\be  I_{TG} = 2 \int\left (B^{IJ}\wedge
    F^{IJ} + \frac{\Lambda}{4}B^{IJ}\wedge B^{IJ}\right) =
    \int \epsilon^{IJKL}\left(B^{IJ}\wedge F^{*KL}
    + \frac{\Lambda}{4}B^{IJ}\wedge B^{*KL}\right),
                                  \lb{6r} \ee
where $F^{*IJ} = \frac 12 \epsilon^{IJKL} F^{KL}$ is the dual
tensor. Here we have the topological sector only in the flat
space, and the dual tensor $F^{*IJ}$ corresponds only to the flat
indexes $I,J$.

The antisymmetric tensor $F^{IJ}$ can be split into a self-dual
component $F^+$ and an anti-self-dual component $F^-$, according
to the relation:
\be
           F^{\pm} = \frac 12 (F \pm F^*).
                                 \lb{6rr} \ee

Here $B^{IJ}$ and $F^{IJ}$ are the following 2-forms:
\be
      B^{IJ} = \theta^I\wedge \theta^J = \frac 12
      \theta_{\mu}^I\theta_{\nu}^Jdx^{\mu}\wedge dx^{\nu} ,
                    \lb{7r} \ee
\be
      F^{IJ} = \frac 12 F_{\mu\nu}^{IJ}dx^{\mu}\wedge dx^{\nu}.
                    \lb{8r} \ee
The tensor $F_{\mu\nu}^{IJ}$ is the field strength (\ref{23}) of
the spin connection $A_{\mu}^{IJ}$.

As it was shown in Refs.~\ct{1a,2a,3a,4a,5a,6a,7a}, actions
$I_{GR}$ and $I_{TG}$, given by Eqs.~(\ref{5r}) and (\ref{6r}),
respectively, can be presented in terms of the self-dual
"left-handed" and anti-self-dual "right-handed" gravity
\ct{7a,9a}:
\be  I_{GR} =  \int [\Sigma^i\wedge F^i - {\bar{\Sigma}}^i\wedge
{\bar F}^i + \Lambda(\Sigma^i\wedge \Sigma^i -
{\bar{\Sigma}}^i\wedge {\bar{\Sigma^i}})],
                      \lb{9r} \ee
and
\be  I_{TG} =  \int [\Sigma^i\wedge F^i + {\bar{\Sigma}}^i\wedge
{\bar F}^i + \Lambda(\Sigma^i\wedge \Sigma^i +
{\bar{\Sigma}}^i\wedge {\bar{\Sigma^i}})],
                      \lb{10r} \ee
where $F\equiv F^+,\quad \bar F\equiv F^-,\quad \Sigma\equiv
\Sigma^+, \quad \bar{\Sigma}\equiv
 \Sigma^-,$
and the left-handed ("+") and right-handed ("--") $F^{\pm
i}_{\mu\nu}$ and $\Sigma^{\pm i}$ are given by:
\be F^{\pm i}_{\mu\nu} =
\partial_{\mu}A^{\pm i}_{\nu} -
\partial_{\nu}A^{\pm i}_{\mu} + \epsilon^{ijk}A^{\pm j}_{\mu}A^{\pm k}_{\nu},
\lb{11r} \ee
\be
    - \Sigma^{\pm i} = - i \theta^0\wedge \theta^i \pm \frac 12
     \epsilon^{ijk}\theta^j\wedge \theta^k.
                          \lb{12r} \ee
Choosing the correct gauge in his gravitational actions, Plebansky
introduced  the Lagrange multipliers $\psi_{ij}$, which are
considered in theory as an auxiliary fields, symmetric and
traceless.

Finally, the resulting actions of the Plebanski gravity are
\ct{1a,2a,3a,4a,5a,6a}:
\be  I_{GR} =  \int [\Sigma^i\wedge F^i - {\bar{\Sigma}}^i\wedge
{\bar F}^i + (\Psi^{-1})_{ij}(\Sigma^i\wedge \Sigma^j -
{\bar{\Sigma}}^i\wedge {\bar{\Sigma^j}})],
                      \lb{13r} \ee
and
\be  I_{TG} = \int [\Sigma^i\wedge F^i + {\bar{\Sigma}}^i\wedge
{\bar F}^i + (\Psi^{-1})_{ij}(\Sigma^i\wedge \Sigma^j +
{\bar{\Sigma}}^i\wedge {\bar{\Sigma^j}})],
                      \lb{14r} \ee
where
\be   (\Psi^{-1})_{ij} = \Lambda\delta_{ij} + \psi_{ij}.
                     \lb{15r} \ee
The explanation of the introduction of the auxiliary field
$\psi_{ij}$ is connected with the condition for metricity. The
'area metric' (see \ct{9a}) is:
\be    m = \frac 12 B^{IJ}\otimes B^{IJ},      \lb{49} \ee
which can be written in terms of $\Sigma^{\pm i}$:
\be
   m^{\pm} = \Sigma^{\pm i}\otimes \Sigma^{\pm i} - \frac 14 V^{\pm}
                       \lb{50} \ee
with
\be  V^{\pm} = \pm 4\Sigma^{\pm i}\wedge \Sigma^{\pm i},
 \lb{51} \ee
where $ V^{\pm}$ both are equal to the 4-volume form $V= \theta^0
\wedge \theta^1 \wedge \theta^2 \wedge \theta^3$. Here $m^+$ and
$V^+$ define a "left" geometry and $m^-$ and $V^-$ define a
"right" geometry of spacetime. These geometries are defined, but
they can be in disagreement. The stationarity with respect to
$\psi_{ij}$ requires the equality $m^+=m^-$ of the left and right
area metrics $m^{\pm}$, which have no totally antisymmetric part,
and leaves the totally antisymmetric parts of $\Sigma^{\pm
i}\otimes \Sigma^{\pm i}$, which form $V$, unconstrained. We see
that $m^+ - m^-$ has 20 linearly independent components. The
equality $m^+=m^-$ provides the correct number of constraints,
reducing the 36 degrees of freedom of ($\Sigma^+,\Sigma^-$) to the
16 degrees of freedom of tetrads $\theta_{\mu}^I$.

The main assumption of Plebanski was that our world, in which we
live, is a self-dual left-handed gravitational world described by
the action:
\be
      I_{(selfdual\,\, GR)}(\Sigma,A) =  \int [\Sigma^i\wedge F^i +
 (\Psi^{-1})_{ij}\Sigma^i\wedge \Sigma^j] ,
                      \lb{52} \ee
and the anti-self-dual right-handed gravitational world is absent
in Nature ($\bar F=0$ and $\bar \Sigma = 0$), and we have the
equality \ct{7a}:
\be     I_{GR} = I_{TG},
 \lb{18r} \ee
i.e. the gravitational sector of gravity coincides with
topological phase of gravity.

Postulating the existence of the Mirror (or Hidden) World
\ct{11,12,13,14,15,16}, we assumed in Ref.~\ct{7a}, that the
anti-self-dual right-handed gravity is the "mirror gravity" given
by the equation:
\be
 I_{(antiselfdual\,\, GR)}(\bar \Sigma, \bar A) = \int [{\bar{\Sigma}}^i\wedge {\bar F}^i
 + (\Psi^{-1})_{ij}{\bar{\Sigma}}^i\wedge {\bar{\Sigma^j}}].
                      \lb{53} \ee
This "mirror gravity" describes the gravity in the Mirror World.

\section{Torsion}

The gravitational theory with torsion can be presented by the
following integral:
\be I_S = 2 \int (2 S^I\wedge S^I + \frac{\Lambda}{4}B^{IJ}\wedge
B^{IJ}),  \lb{54} \ee
where the 2-form
\be  S^I =\frac 12 S_{\mu\nu}^Idx^{\mu}\wedge dx^{\nu}
            \lb{19r} \ee
contains the torsion $S_{\mu\nu}^I$, which is the field strength
(\ref{25}) of the tetrads $\theta_{\mu}^I$.

Using the partial integration and putting
\be \int \partial_{\mu}(T_{\nu\kappa\lambda})dx^{\mu}\wedge
dx^{\nu}\wedge dx^{\kappa}\wedge dx^{\lambda}=0,
 \lb{16x} \ee
it is not difficult to show that
\be \int B^{IJ}\wedge
    F^{IJ} = 2\int S^I\wedge S^I.  \lb{17x} \ee
According to Eqs.~(\ref{6r}) and (\ref{54}), we have:
\be
 I_{TG} = 2 \int \left(B^{IJ}\wedge
    F^{IJ} + \frac{\Lambda}{4}B^{IJ}\wedge B^{IJ}\right) =
2 \int \left(2 S^I\wedge S^I + \frac{\Lambda}{4}B^{IJ}\wedge
B^{IJ}\right) = I_S. \lb{18x} \ee
This means that in the self-dual Plebanski formulation of
gravitational theory (i.e. in our world) all sectors of gravity
coincide \ct{7a}:
\be
  I_{GR} = I_{TG} = I_S.
 \lb{55} \ee
Here we see a close analogy of the  geometry of spacetime in GR
with a structure of defects in a crystal \ct{1,2,3}. Torsion is
presented by the translational defects --dislocations, and
curvature is given by the rotational defects -- disclinations. But
these defects are not independent of each other: a dislocation is
equivalent to a disclination-antidisclination pair, and a
disclination gives a string of dislocations.

The crystalline defects present a special version of the curved
spacetime - the Riemann-Cartan spacetime which has an additional
geometric property - "torsion" \ct{4}, described by dislocations.
The world crystal is a model for the Einstein's gravitation which
has a new type of gauge symmetry with a zero torsion as a special
gauge, while a zero connection (with zero Cartan curvature) is
another equivalent gauge with nonzero torsion which corresponds to
the Einstein's theory of "teleparallelism" \ct{5}. But any
intermediate choice of the gauge for the field $A^{IJ}_{\mu}$ is
also allowed \ct{3}. In the present investigation we have shown
that in the Plebanski formulation the phase of gravity with
torsion is equivalent to the ordinary or topological gravity, and
we can exclude a torsion as a separate dynamical variable. This
explains why the Einstein's theory of gravity described only by
curvature corresponding to the disclination defects of a crystal
can be rewritten as a teleparallel theory of gravity described
only by torsion which corresponds to the dislocation defects.

\section{Summary and Conclusions}

In this paper we have shown the close analogy of geometry of
space-time in GR with a structure of defects in a crystal, given
by H.~Kleinert in his model \ct{1,2,3}. This model was related
with the Plebanski formulation of gravity. In the Section 1 we
have discussed a role of torsion. We have reviewed the Kleinert's
model of a crystal with defects in the Section 2 considering the
translational defects -- dislocations, and the rotational defects
-- disclinations -- in the 3- and 4-dimensional crystals. In the
Section 3 we have explained the main idea of Plebanski \ct{1a} to
construct the 4-dimensional gravitational action without metric.
The integrand of this action contains a product of two 2-forms,
which are constructed from the tetrads $\theta^I$ and the
connection $A^{IJ}$ considered as independent dynamical variables.
Here we have presented the relation between the Kleinert's model
of a crystal with defects and Plebanski's formulation of gravity.
It was shown that the tetrads $\theta^I_{\mu}$ are dislocation
gauge fields, and the connections $A^{IJ}_{\mu}$ are the
disclination gauge fields, the curvature is the field strength of
connections $A^{IJ}_{\mu}$, and the torsion is the field strength
of tetrads $\theta^I_{\mu}$. Both $A^{IJ}$ and $\theta^I$ are
1-forms. The tetrads $\theta^I_{\mu}$ were used instead of the
metric $g_{\mu\nu}$.

Section 4 is devoted to the torsion. The crystalline defects
represent a special version of the curved space-time -- the
Riemann--Cartan space-time with torsion \ct{4}. The world crystal
gives a model with a new type of gauge symmetry for GR \ct{3}. The
Einstein's gravitation, which has a zero torsion, corresponds to a
special gauge, while the theory with a zero connection (i.e. zero
Cartan curvature) is another equivalent gauge with nonzero torsion
which corresponds to the Einstein's theory of "teleparallelism"
\ct{5}. In this paper we have showed that in the Plebanski
formulation of gravity the torsional gravitational phase (phase
described only by nonzero torsion) is equivalent to the ordinary
or topological gravity, and we can exclude torsion as a separate
dynamical variable of gravity. In the Plebanski formulation of
4-dimensional gravity the Einstein's gravity with a space-time
corresponding to the structure of a crystal with disclinations,
i.e. rotational defects, coincides with the Einstein's
"teleparallel" gravitational theory \ct{5} corresponding to a
crystal having only the translational defects -- dislocations.

\section{Acknowledgments}

H.B.N. thanks for being alluded hospitality and office to finish
this work. L.V.L. thanks the Niels Bohr Institute for the
wonderful hospitality and financial support. The authors deeply
thank M.~Chaichian and A.~Tureanu from Helsinki University for
useful discussions and help. CRD acknowledges a scholarship from 
the Funda\c c\~ ao para a Ci\^ encia e a Tecnologia (FCT, Portugal) 
(ref.~SFRH/BPD/41091/2007) and this work was 
partially supported by the FCT projects CERN/FP/123580/2011,
PTDC/FIS/098188/2008 and CFTP-FCT Unit 777 which are partially funded
through POCTI (FEDER).

\end{document}